\documentclass{article}
\usepackage{multind}
\usepackage[T1]{fontenc}
\usepackage{ae,aecompl}
\usepackage[utf8]{inputenc}
\usepackage[english]{babel}
\usepackage{amsmath}
\usepackage{amsfonts}
\usepackage{amssymb}
\usepackage{amsthm}
\usepackage{times}
\usepackage{verbatim, amsbsy}
\usepackage{graphicx}
\usepackage{epsfig}
\usepackage[normalem]{ulem}
\usepackage{color}
\usepackage{fancyhdr}
\usepackage{lastpage}
\usepackage{listings}
\usepackage{transparent}
\usepackage{bm}
\usepackage{subfigure}
\usepackage[all]{xy}
\usepackage{url}
\usepackage{import}
\usepackage{spconf}
\usepackage{extarrows}

\newfont{\bbb}{msbm10 scaled 500}

\newfont{\bb}{msbm10 scaled 1100}

\newcommand{\hv}{{\bf h}}

\newcommand{\nv}{{\bf n}}

\newcommand{\sv}{{\bf s}}

\newcommand{\xv}{{\bf x}}
\newcommand{\yv}{{\bf y}}

\newcommand{\zerov}{{\bf 0}}

\newcommand{\Am}{{\bf A}}
\newcommand{\Bm}{{\bf B}}

\newcommand{\Em}{{\bf E}}
\newcommand{\Fm}{{\bf F}}
\newcommand{\Gm}{{\bf G}}
\newcommand{\Hm}{{\bf H}}
\newcommand{\Id}{{\bf I}}

\newcommand{\Km}{{\bf K}}

\newcommand{\Pm}{{\bf P}}

\newcommand{\Sm}{{\bf S}}

\newcommand{\Um}{{\bf U}}

\newcommand{\Vm}{{\bf V}}

\newcommand{\Zm}{{\bf Z}}

\newcommand{\Ac}{{\cal A}}
\newcommand{\Bc}{{\cal B}}
\newcommand{\Cc}{{\cal C}}

\newcommand{\Nc}{{\cal N}}

\newcommand{\etav}{\hbox{\boldmath$\eta$}}

\newcommand{\Gammam}{\hbox{\boldmath$\Gamma$}}
\newcommand{\Lambdam}{\hbox{\boldmath$\Lambda$}}

\newcommand{\Sigmam}{\hbox{\boldmath$\Sigma$}}

\newcommand{\trace}{{\hbox{tr}}}

\newcommand{\herm}{^{\text{H}}}

\newcommand{\ax}{{\text{a}}}
\newcommand{\bx}{{\text{b}}}

\newcommand{\gx}{{\text{g}}}

\newcommand{\px}{{\text{p}}}

\newcommand{\sx}{{\text{s}}}

\newcommand{\executeiffilenewer}[3]{%
\ifnum\pdfstrcmp{\pdffilemoddate{#1}}%
{\pdffilemoddate{#2}}>0%
{\immediate\write18{#3}}\fi%
}

\DeclareMathOperator{\rank}{\mathrm{rank}}

\DeclareMathOperator{\dimV}{\mathrm{dim}}
\DeclareMathOperator{\spanV}{\mathrm{span}}

\newtheorem{lemma}{Lemma}

\newtheorem{theorem}{Theorem}
\newtheorem{definition}[theorem]{Definition}

\newenvironment{thmproof}{{\bf Proof:}}{\hfill\rule{2mm}{2mm}}

\begin{document}

\title{Cognitive Interference Alignment for OFDM two-tiered Networks}
\name{Marco Maso$^{\star,\diamond}$, Leonardo S. Cardoso$^{\dagger}$, M\'erouane Debbah$^\star$ and Lorenzo Vangelista$^\diamond$}
\address{$\star$ Alcatel-Lucent Chair - SUP\'ELEC, Gif-sur-Yvette, France\\ $\diamond$ Department of Information Engineering, University of Padova, Italy\\ $\dagger $ INSA - Lyon, France} 
   
\maketitle

\begin{abstract}
In this contribution, we introduce an interference alignment scheme that allows the coexistence of an orthogonal frequency division multiplexing (OFDM) macro-cell and a cognitive small-cell, deployed in a two-tiered structure and transmitting over the same bandwidth. We derive the optimal linear strategy for the single antenna secondary base station, maximizing the spectral efficiency of the opportunistic link, accounting for both signal sub-space structure and power loading strategy. Our analytical and numerical findings prove that the precoder structure proposed is optimal for the considered scenario in the face of Rayleigh and exponential decaying channels.
\end{abstract}

\vspace{-.3cm}
\section{Introduction}

In recent years, new technologies have been studied to overcome the capacity shortfall and the ever-present coverage issue of current 3 and 3.5G networks. One solution is the, so-called heterogeneous tiered networks, composed of macro-cells and small-cells that coexist in the same coverage area. One of the biggest challenges two-tiered networks face is the definition of a suitable strategy to realize the coexistence of the two tiers in a spectrum sharing approach. On the one hand, if the two tiers communicate over the same bandwidth the overall spectral efficiency increases, while on the other hand, high levels of interference are generated. 

Joint beamforming \cite{art:bhaga11} or cooperative frequency reuse \cite{conf:akoum10} strategies can be implemented to deal with both cross- and co-tier interference if the two tiers cooperate. When no cooperation is possible (or desirable), techniques based on dynamic spectrum access (DSA) and cognitive radio \cite{art:goldsmith2009bs} are a solution, due to their capability to adapt intelligently to the environment. One DSA technique, called Vandermonde-subspace frequency division multiplexing (VFDM) \cite{conf:cardoso2010v}, uses the frequency dimension and perfect channel state information (CSI) to allow the coexistence of macro- and small-cells. This is done through the use of a null-space precoder that protects the macro- from small-cells transmissions. Interference alignment (IA) \cite{art:cadambe2008i} can also be used if CSI is globally available in the network. In particular, if multiple spatial dimensions are available at one of the tiers, i.e. for multiple input multiple output (MIMO) systems, IA can provide an effective cross-tier interference management \cite{conf:wu10}. Additionally, if the cognitive tier is aware of the power allocation in the primary tier, solutions such as opportunistic interference alignment \cite{art:perlaza2009, conf:ganesan09} can be adopted to for the coexistence.  

In this contribution, we focus on a two-tiered network, whose licensee macro-cell base station (MBS) is OFDM based and oblivious of a cognitive secondary base station (SBS), transmitting over the same bandwidth without cooperation. Herein, single input single output (SISO) transmissions are performed in both tiers. Furthermore, the SBS is uninformed about left-over space, time or frequency resources or power allocation in the macro-cell. Inspired from VFDM, a novel overlay cognitive interference alignment (CIA) scheme is proposed to increase the spectral efficiency of the two-tiered network. The optimal linear strategy that maximizes the spectral efficiency of the secondary link is derived, accounting for both signal sub-space structure and power loading strategy. We show that, for Rayleigh uniform power delay profile (PDP) channels, VFDM falls into the CIA semi-unitary precoder case, and therefore, is optimal. Conversely, for exponentially decaying PDP channels, we show how CIA outperforms the root-based precoder structure, demonstrating a consistent optimality of the performance.  

This paper is organized as follows. In Sec. \ref{sec:model}, we introduce the considered model. In Sec. \ref{sec:cia}, we describe CIA. We derive the optimal precoder in Sec. \ref{sec:precoder_c}. In Sec. \ref{sec:num_anal}, we provide a numerical analysis that supports our claims. Finally, conclusions and future research directions are discussed in Sec.~\ref{sec:conclusion}.

\vspace{-.4cm}
\section{Model} \label{sec:model}

\begin{figure}[!h]
\begin{center}
\includegraphics[width=\linewidth]{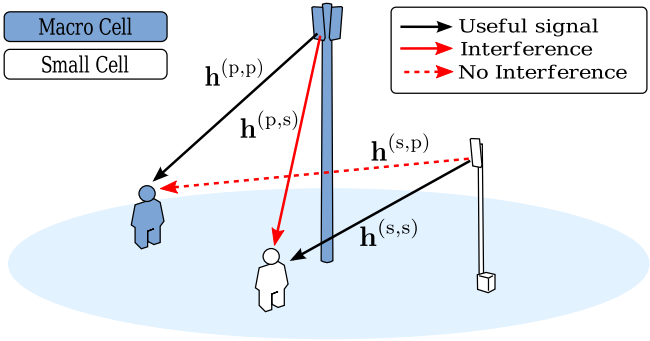} % requires the graphicx package
        \vspace{-.4cm}
        \caption{A downlink two-tiered network}
        \label{fig:scenario}
\end{center}
\end{figure}

Consider the downlink of the two-tiered network depicted in Fig. \ref{fig:scenario}, where a single antenna cognitive SBS is deployed inside the coverage area of a licensee single antenna MBS. For simplicity, we assume that both transmitters serve a single user equipment. The MBS is an OFDM transmitter, as in recent standards (e.g. \cite{rpt:3gpp25.814}). Both macro and secondary user equipments (MUE/SUE) are classic OFDM receivers. Both tiers are independent and transmit over the same bandwidth with no cooperation or coordination strategy between tiers. The legacy system is unaware of the existence of the opportunistic one, and does not implement any interference management scheme. This model is equivalent to the cognitive interference channel (CIC) model, but with no cooperation between systems.

In the notation used throughout the work, the primary system is referred by the subscript ``$\px$'' and the secondary by ``$\sx$''. $\Id_M$ is the $M \times M$ identity matrix and $\zerov_{N \times M}$, the $N \times M$ zeros matrix. Let $\hv^{(\px,\px)}$, $\hv^{(\px,\sx)}$, $\hv^{(\sx,\px)}$, $\hv^{(\sx,\sx)} \sim {\Cc}\Nc(0,\Id_{l+1}/(l+1))$ be i.i.d. Rayleigh fading channel vectors of $l+1$ taps. For simplicity, but without lack of generality, we consider that all the channel vectors have the same size. Moreover, we assume that both systems adopts Gaussian constellations. Let $N$ be number of the subcarriers used by the MBS, and $L \geq l$ the cyclic prefix length to compensate for the inter-symbol (ISI) and inter-block interference (IBI). For a transmitter ``$\ax$'' and receiver ``$\bx$'', the channel convolution matrix $\Hm_{\ax \bx} \in \Cc^{(N+L)\times (N+L)}$ is defined as 
\begin{equation*}
\begin{footnotesize}
\Hm_{\ax\bx} =
\left[ \begin{array}{ccccccc}
h_0^{(\ax,\bx)} & 0 & \cdots & h_{l}^{(\ax,\bx)} & \cdots & \cdots & h_1^{(\ax,\bx)} \\
\vdots & \ddots & \ddots &  & \ddots & \ddots & \vdots \\
\vdots & \ddots & \ddots &  & \ddots & \ddots & h_{l}^{(\ax,\bx)} \\  
h_{l}^{(\ax,\bx)} & \cdots & \cdots & h_0^{(\ax,\bx)} & 0 & \cdots & 0\\
0 & \ddots & \ddots & \ddots & \ddots & \ddots & \vdots \\
\vdots & \ddots & \ddots & \ddots & \ddots & \ddots & 0 \\ 
0 & \ddots & \ddots & h_{l}^{(\ax,\bx)} & \ddots & \ddots & h_0^{(\ax,\bx)}\\ 
 \end{array} \right],
 \end{footnotesize}
\end{equation*}
considering that the channel coherence time is largely superior to the block transmission time $N+L$ such that the channel is essentially the same from one block to the other.

Let $\xv_\px$, $\xv_\sx  \in \Cc^{(N+L) \times 1}$ be the transmit symbol vectors and $\yv_\px$, $\yv_\sx  \in \Cc^{(N+L) \times 1}$ be the received symbol vectors for the primary/secondary system respectively. The overall received signal at primary and secondary receiver is
\begin{eqnarray}\label{eq:originaly1}
        %\begin{cases}
    \yv_{\px} = \Hm_{\px\px}\xv_{\px} + \Hm_{\sx\px}\xv_{\sx} + \nv_{\px} \\
    \yv_{\sx} = \Hm_{\sx\sx}\xv_{\sx} + \Hm_{\px\sx}\xv_{\px} + \nv_{\sx},\nonumber     
    %\end{cases}
\end{eqnarray}
where $\nv_\px$, $\nv_\sx \sim\Cc\Nc(0, \sigma^2 \Id_{N+L})$ are $N+L$-sized AWGN noise vectors. Note that, this model could be easily extended to the multi-user case at the primary and secondary system, thus, the single MUE/SUE hypothesis does not affect its generality. The OFDM transmitted signal by the MBS can be expressed as
\begin{equation}
\xv_{\px} = \Am \Fm^{-1} \sv_\px,
\end{equation}
where $\Am$ is a $(N+L)\times N$ cyclic prefix precoding matrix, $\Fm \in \Cc^{N\times N}$ is a unitary DFT matrix and $\sv_{\px}$ is a zero mean, unit norm primary input symbol vector. The SBS precodes its signal with an appropriate pre-processing matrix $\Em$, such that
\begin{equation} \label{eq:s_inp_vec}
   \xv_\sx = \Em \sv_\sx,
\end{equation}
with $\Em$ and $\sv_\sx$ detailed in the following. Note that, when a perfect CSI at the opportunistic transmitter is available, the state-of-the-art solution for the considered scenario is given by a precoder $\Em$, devised according to the VFDM orthonormal root-based scheme we proposed in \cite{conf:cardoso2010v}. In this work, we take a step further and seek for the optimal linear strategy to maximize the spectral efficiency of the secondary link, while nulling the cross-tier interference toward the MUE. Therefore, we frame the coexistence problem in an IA perspective, providing a signal subspace based approach characterized by a higher analytic tractability. 

\vspace{-.2cm}
\section{Cognitive interference alignment} \label{sec:cia}

We assume that the cognitive SBS disposes of a perfect CSI w.r.t. $\hv^{(\sx,\px)}$ and $\hv^{(\sx,\sx)}$. On the other hand, it does not adopt spectrum sensing techniques nor has any a priori information about the time resource allocation in the primary system. Moreover, it does not have knowledge about the primary transmit input symbol vector, thus techniques such as dirty paper coding (DPC) \cite{art:costa83} can not be implemented. 

Let us consider the classic OFDM receiver chain, where the following baseband pre-processing is performed
\begin{eqnarray} \label{eq:y_tilde}
\tilde{\yv}_{\px} &=& \Fm \Bm \yv_\px = \tilde{\Hm}_{\px\px} \Am \Fm^{-1} \sv_\px + \tilde{\Hm}_{\sx\px} \xv_\sx + \tilde{\mathbf{\nv}}_{\px} \\
\tilde{\yv}_{\sx} &=& \Fm \Bm \yv_\sx = \tilde{\Hm}_{\px\sx} \Am \Fm^{-1} \sv_\px + \tilde{\Hm}_{\sx\sx} \xv_\sx + \tilde{\mathbf{\nv}}_{\sx},\nonumber
\end{eqnarray}
where $\Bm=\left[\zerov_{N \times L} | \Id_N\right]$ is the cyclic prefix removal matrix and $\tilde{\mathbf{\nv}}_{\sx}$, $\tilde{\mathbf{\nv}}_{\px}$ are the Fourier transform of the last $N$ elements of the noise vectors $\nv_{\sx}$, $\nv_{\px}$, having the same statistics. In particular $\tilde{\Hm}_{\ax\bx} \in \Cc^{N \times (N+L)} = \Fm\Bm\Hm_{\ax\bx}$ reads
\begin{eqnarray*}
\begin{footnotesize}
\tilde{\Hm}_{\ax\bx} =
\Fm \left[ \begin{array}{cccccc}
h_{l}^{(\ax,\bx)} & \cdots & h_0^{(\ax,\bx)} & 0 & \cdots & 0 \\
0 & \ddots &  & \ddots & \ddots & \vdots \\
\vdots & \ddots & \ddots &  & \ddots & 0 \\
0 & \cdots & 0 & h_{l}^{(\ax,\bx)} & \cdots & h_0^{(\ax,\bx)} \\
\end{array} \right],
\end{footnotesize}
\end{eqnarray*}
and $\rank({\tilde{\Hm}_{\ax\bx}}) = N$. From the rank-nullity theorem we have
\begin{equation}
 \dimV{\ker({\tilde{\Hm}_{\sx\px}})}= L,
\end{equation}
$\forall \hv^{(\sx,\px)} \in \Cc^{(l+1) \times 1}$. Therefore, we can always find a matrix $\Em \in \Cc^{(N+L) \times L}$ such that 
\begin{equation} \label{eq:nullcondition}
    \spanV{(\Em)} = \ker({\tilde{\Hm}_{\sx\px}})\footnote{Let $\Ac$ and $\Bc$ be two vector spaces of dimension $M$. We define $\Ac = \Bc$ if and only if $\forall x \in \Cc^{M}$, $x \in \Ac \leftrightarrow x \in \Bc$.}
\end{equation}
and $\tilde{\Hm}_{\sx\px}\Em = \zerov_{N \times L}$. At this stage, we can let $\sv_{\sx} \in ~\Cc^{L \times 1}$ in \eqref{eq:s_inp_vec} be the zero mean unit norm secondary input symbol vector.

According to the IA paradigm, the signal coming from the secondary system must be confined in a constant sized subspace of the overall received signal space. By substituting \eqref{eq:s_inp_vec} in \eqref{eq:originaly1}, we obtain 
\vspace{-.1cm}
\begin{equation} \label{eq:y_intal}
\yv_{\px} = \Hm_{\px\px}\xv_{\px} + \left[\begin{array}{c}\Km \\ \zerov_{N \times L} \end{array} \right] \sv_{\sx} + \nv_{\px},  \\ 
\end{equation}
where $\Km \in \Cc^{L \times L}$ is a matrix whose size is independent of the size of $\hv^{(\sx,\px)}$, i.e., $l$. At this stage, the MUE can obliviously extract the desired $N$ interference free dimensions out of the $N+L$ received ones thanks to the OFDM receiver pre-processing shown in \eqref{eq:y_tilde}, eliminating the aligned interference coming from the SBS. This gives us
\begin{equation} \label{eq:y_tilde2}
\tilde{\yv}_\px = \tilde{\Hm}_{\px\px} \xv_\px + \tilde{\nv}_{\px}.
\end{equation}

\vspace{-.5cm}
\section{Optimal CIA precoder} \label{sec:precoder_c}

As seen in Sec. \ref{sec:model}, the adoption of the CIA scheme preserves the degrees of freedom of the primary OFDM transmission, hence its maximum achievable spectral efficiency. On the other hand, thanks to the joint effect of the receiver pre-processing and the redundancy introduced by the MBS to combat ISI and IBI, the CIA scheme guarantees the SBS counts with $L$ additional transmit dimensions. Naturally, the efficiency of the secondary transmission hinges on the choice of the precoder $\Em$, that has to be designed such that the spectral efficiency of the secondary link is maximized. Let us start from a definition.
\vspace{-.1cm}
\begin{definition}[Semi-unitary precoder]
A precoder $\Zm \in \Cc^{N \times M}$ is \textit{semi-unitary} if and only if $\rank{(\Zm)}=min\{N,M\}$ and all its non zero eigenvalues are equal to 1, thus $\Zm\Zm\herm=\Id_N$ or $\Zm\herm\Zm=\Id_M$.
\end{definition}
The optimal choice for the SBS to design a spectral efficiency maximizing precoder is given by the following result.
\vspace{-.3cm}
\begin{lemma}[Optimal CIA precoder] \label{lemma:optimal}
Consider the single antenna CIC model, where the messages $\sv_{\px}$ and $\sv_{\sx}$ to be transmitted are known solely at their respective transmitters. The spectral efficiency of the secondary link is maximized by a semi-unitary precoder structure and water-filling power loading strategy. 
\end{lemma}

\vspace{-.1cm}
\begin{thmproof}
Let us start with some preliminary considerations. The MBS adopts a uniform power allocation as in \cite{rpt:3gpp25.814}, with a power of $P_\px$ per input symbol. There is no cooperation between the two tiers, hence, no information about the primary system's message is available at the secondary system. In particular, the SUE performs single-user decoding, i.e., interference is seen as noise. Let $\etav = \tilde{\Hm}_{\px\sx}\Am \Fm^{-1} \sv_\px + \tilde{\mathbf{\nv}}_{\sx}$ be the interference plus noise component of the received message at the SUE. The received signal at the SUE becomes 
\begin{displaymath}
\tilde{\yv}_{\sx} = \tilde{\Hm}_{\sx\sx} \Em \sv_{\sx} + \etav.
\end{displaymath}
Let $\Sm_{\sx}=\Em\Pm\Em\herm$ be the SBS input covariance matrix, where $\Pm=diag[p_1, ..., p_L]$ is the covariance matrix of $\sv_\sx$. We can approximate $\etav$ to a zero-mean Gaussian random vector with covariance matrix given by $\Sm_{\eta} =  \tilde{\Hm}_{\px\sx} \Sm_{\px} \tilde{\Hm}_{\px\sx}\herm + \sigma_n^2 \Id_N$, where $\Sm_{\px}=P_\px\Fm\Am^{\text{T}}\Am\Fm^{-1}$ is the transmit covariance matrix at the MBS. At this stage, we assume perfect knowledge of $\Sm_{\eta}$ at the SBS. 

Let $P_{\sx}$ be the transmit power per precoded symbol. Then, the maximum achievable spectral efficiency for the secondary system is the solution of the following maximization problem
\begin{eqnarray}\nonumber \label{eq:max_1}
    \max_{\Sm_\sx} & & \frac{1}{N+L} \log_{2}\left|\Id_N + \Sm_{\eta}^{-1/2}\tilde{\Hm}_{\sx\sx}\Sm_\sx \tilde{\Hm}_{\sx\sx}\herm \Sm_{\eta}^{-1/2}\right|  \\ 
 \text{s.t.}  & &  \tilde{\Hm}_{\sx\sx} \Em = \textbf{0}_{N \times L} \\ \nonumber & &  \trace(\Sm_\sx) \leq (N+L)P_{\sx}.
\end{eqnarray}
As previously stated, the constraint $\tilde{\Hm}_{\sx\px} \Em = \textbf{0}_{N \times L}$ implies that the columns of $\Em$ have to span $\ker{(\tilde{\Hm}_{\sx\px})}$. Its presence restricts the subset of the possible solutions to the kernel of the equivalent interference channel. As a consequence, let $\Vm$ be a matrix whose columns form an orthonormal basis of $\ker{(\tilde{\Hm}_{\sx\px})}$, then we can remove the constraint and write
\begin{equation} \label{eq:precoder}
\Em = \Vm \Gammam,
\end{equation}
where $\Gammam \in \Cc^{L \times L}$ is a complex linear combination matrix. In particular, we remark that the columns of $\Em$ are a generic linear combination of the columns of the basis, hence any optimal $\Em^*=\Vm \Gammam^*$ will always satisfy $\tilde{\Hm}_{\sx\px} \Em^* = \textbf{0}_{N \times L}$. Note that, $\Vm$ is semi-unitary by definition of the orthonormal matrix and several equivalent techniques can be adopted to derive it, e.g. singular value decomposition (SVD). Then
\begin{equation*} \label{eq:sigma}
\Sm_{\sx} =  \Vm \Gammam \Pm \Gammam\herm \Vm\herm = \Vm \Sigmam_\sx \Vm\herm,
\end{equation*}
with $\Sigmam_\sx = \Gammam \Pm \Gammam\herm$, and \eqref{eq:max_1} becomes
\begin{eqnarray*}
\max_{\Sigmam_\sx} & & \frac{1}{N+L} \log_{2}\big|\Id_N + \Sm_{\eta}^{-1/2}\tilde{\Hm}_{\sx\sx}\Vm \Sigmam_\sx \Vm\herm \tilde{\Hm}_{\sx\sx}\herm \Sm_{\eta}^{-1/2}\big| \\ \nonumber
\text{s.t.}&& \hspace{.1cm} \trace(\Sigmam_\sx) \leq (N+L)P_{\sx}.\nonumber
\end{eqnarray*}
Let $\Gm=\Sm_{\eta}^{-1/2}\tilde{\Hm}_{\sx\sx} \Vm$ and $\Gm = \Um_{\gx}\Lambdam_{\gx}^{\frac{1}{2}}\Vm_{\gx}^{\text{H}}$ be its SVD, with $\Um_{\gx}\in\Cc^{N\times N}$,$\Vm_{\gx}\in\Cc^{L\times L}$ unitary matrices. $\Lambdam_{\gx}=~[\Lambdam_{\gx}^{\lambda}, \Lambdam_{\gx}^{0}]^{\text{T}}$, where $\Lambdam_{\gx}^{\lambda}$ is a diagonal matrix carrying the $L$ eigenvalues of $\Gm \Gm^{\text{H}}$ and $\Lambdam_{\gx}^{0}=\textbf{0}_{L \times (N-L)}$. Therefore, we can write
\begin{eqnarray}\nonumber \label{eq:max_2}
    \max_{\Sigmam_\sx} & & \frac{1}{N+L} \log_{2}\left|\Id_N + \Um_{\gx}\Lambdam_{\gx}^{\frac{1}{2}}\Vm_{\gx}\herm \Sigmam_\sx \Vm_{\gx}\Lambdam_{\gx}^{\frac{1}{2}} \Um_{\gx}\herm \right|  \\
 \text{s.t.}  & &   \trace(\Sigmam_\sx) \leq (N+L)P_{\sx} .
\end{eqnarray}
By Hadamard inequality, we know that the determinant of a positive definite matrix is upper-bounded by the product of the elements on its main diagonal, i.e., $|\Am| \leq \prod_i \Am_{[i,i]}$. This implies that, in order to diagonalize the argument of the determinant in \eqref{eq:max_2}, $\Sigmam_\sx = \Gammam \Pm \Gammam\herm = \Vm_\gx \Pm \Vm_\gx\herm$, thus $\Gammam^* = \Vm_\gx$, and \eqref{eq:max_2} becomes
\begin{eqnarray}\label{eq:newproblem}
    \max_{p_i} & & \sum_{i=1}^{L}\log_{2}(1 + p_{i}[\Lambdam_{\gx}^{\lambda}]_{i,i})  \\
    \text{s.t.} & & \sum_{i=1}^{L} p_{i} \leq (N+L)P_{\sx}. \nonumber
\end{eqnarray}
By applying the classical water-filling algorithm, we find the solution to \eqref{eq:max_1}, $\Sm_{\sx} = \Vm \Vm_{\gx} \Pm \Vm_{\gx}\herm \Vm \herm$, where the $i-$th component of the matrix $\Pm$ is the water-filling solution of \eqref{eq:newproblem}
\begin{equation}\label{eq:wf2}
p_i = \left[\mu -\frac{1}{[\Lambdam_{\gx}^{\lambda}]_{i,i}}\right]^+,
\end{equation}
with ``water level'' $\mu$ , determined such as $\sum_i^L p_i \leq (N+~L)P_{\sx}$. By plugging the optimal solution into \eqref{eq:precoder} we get
\begin{equation}
\Em^* = \Vm \Vm_{\gx},
\end{equation}
precoder that maximizes the spectral efficiency under the considered constraints. By definition of SVD, $\Vm_{\gx}$ is unitary, hence $\Em$ is semi-unitary and this ends the proof.
\end{thmproof}

\vspace{-.2cm}
\section{Numerical analysis} \label{sec:num_anal}

In this section we present the results obtained through Monte Carlo simulations of a transmission performed by the SBS. We consider three power delay profile (PDP) models for the aforementioned Rayleigh fading channels, i.e., uniform, exponential with fast ($\frac{T_s}{\tau}=2$) and slow ($\frac{T_s}{\tau}=0.75$) decay, where $T_s$ is the sample time and $\tau$ is the root mean square (r.m.s.) delay spread. We focus on the maximum achievable spectral efficiency of the secondary link, thus we neglect the impact of the primary system interference on the SUE, identified as the subject of our future research. We compare the achievable performance of the CIA unitary precoder and show the gains that this approach can yield w.r.t. non unitary precoders (suboptimal approach according to Lemma \ref{lemma:optimal}). As a further complementary benchmark, we consider a unitary root-based VFDM precoder, derived by means of a Gram-Schmidt orthonormalization \cite{conf:cardoso2010v}, and evaluate its performance for the considered operative scenarios. We adopt for VFDM the same power loading strategy as the optimal CIA solution. We assume that the MBS transmits over $N=128$ subcarriers, with a cyclic prefix size of $L=32$ and that the channel size $l$ coincides with the cyclic prefix size $L$. 
 
\begin{figure}[!h]
	\centering
	\includegraphics[width=\linewidth]{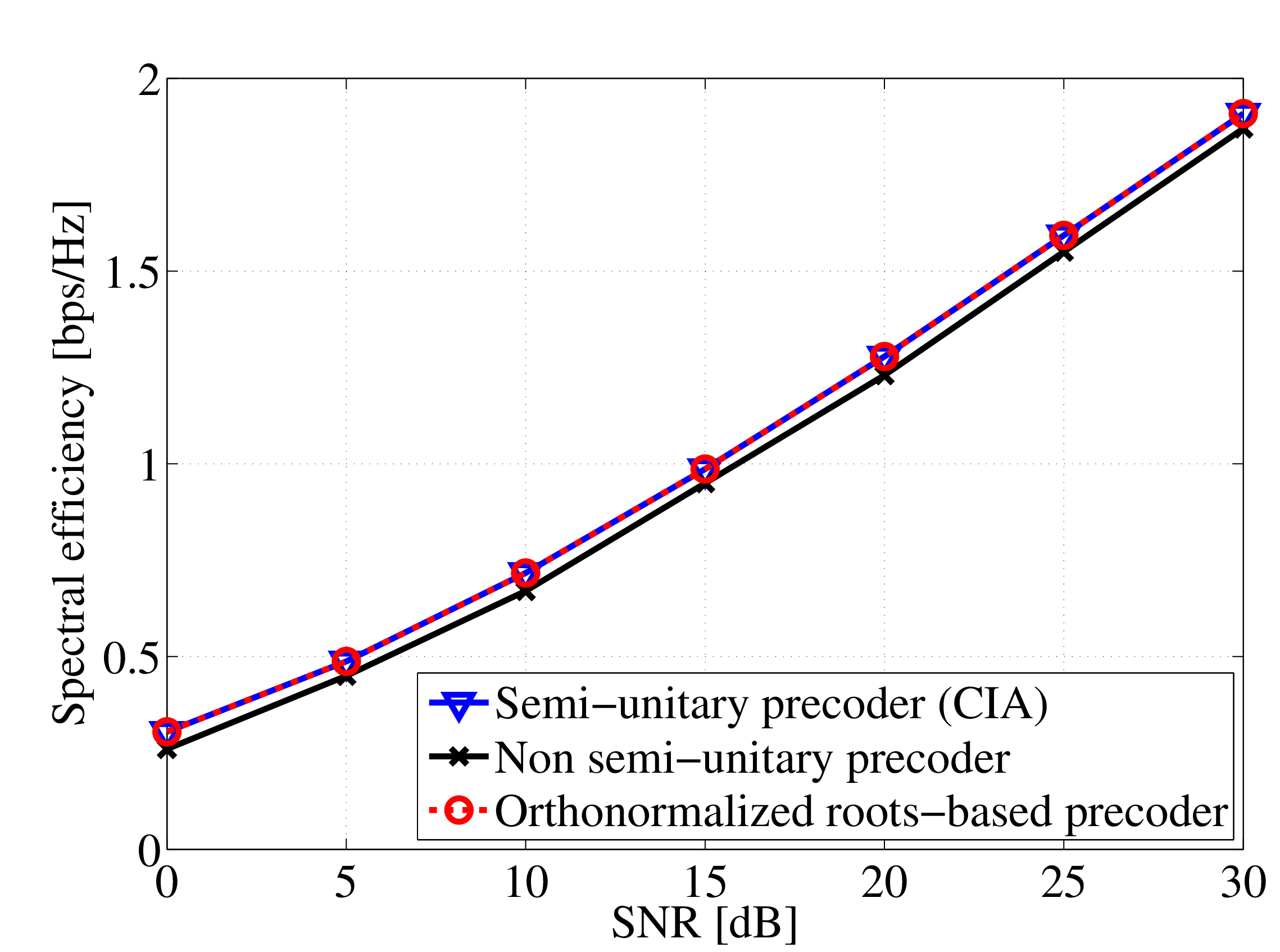} % requires the graphicx package
	\vspace{-0.8cm}
	\caption{Spectral efficiency of the secondary link. Uniform PDP.}
	 \label{fig:uniform}
\end{figure}

We start from the uniform PDP, in Fig. \ref{fig:uniform}. 
%
%\vspace{-.2cm}

If compared to the optimal performance provided by the semi-unitary (CIA) precoder, we notice that, by adopting a sub-optimal solution, a loss of less than $3\%$ can be seen for the considered SNR range. On the other hand, the performance of the unitary root-based VFDM and CIA precoder are identical. This demonstrates the optimality of the results provided in \cite{conf:cardoso2010v} when the considered channels are characterized by a uniform PDP.

\begin{figure}[!h]
	\centering
	\includegraphics[width=\linewidth]{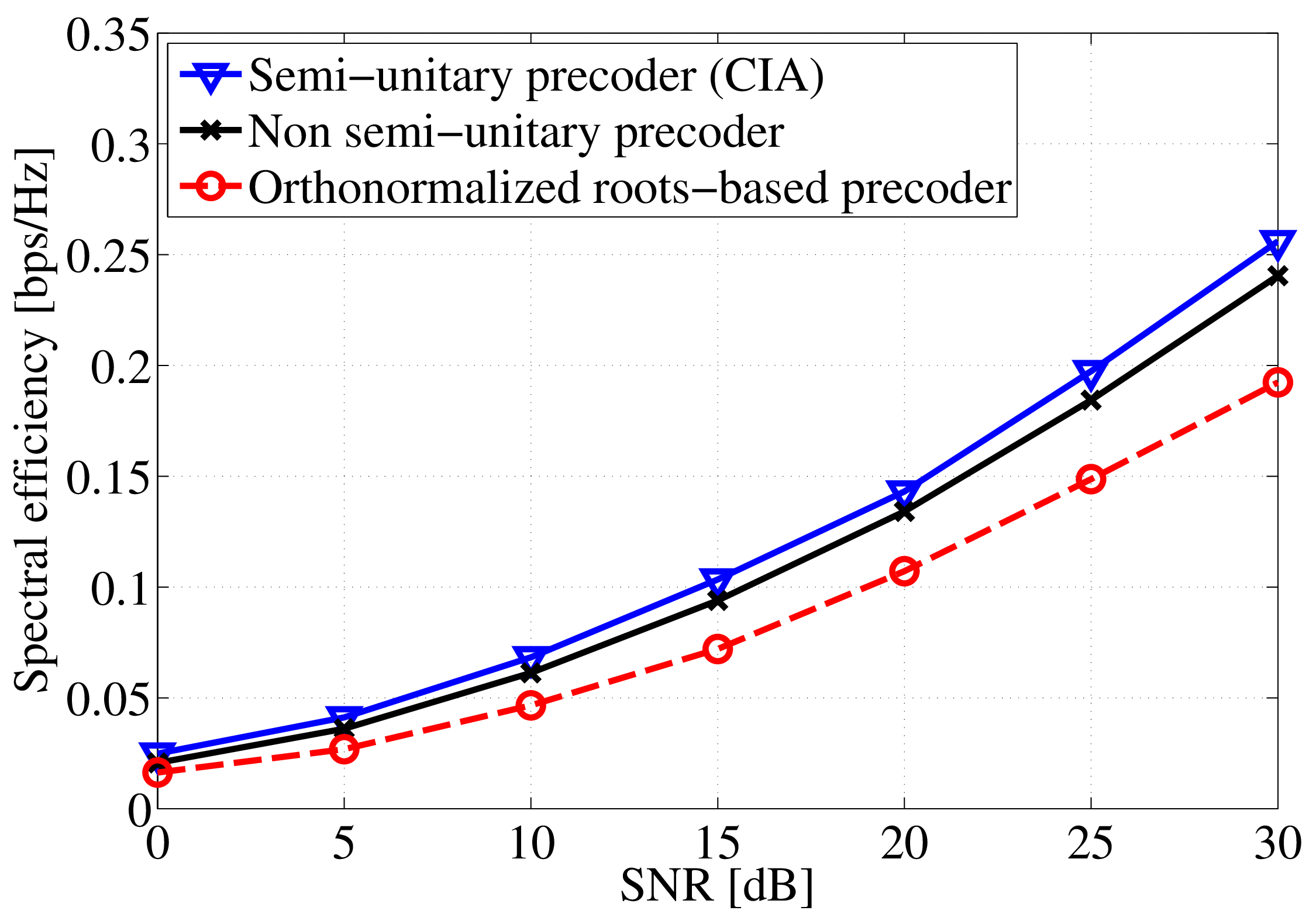} % requires the graphicx package
	\vspace{-0.8cm}
	\caption{Spectral efficiency of the secondary link. Exponential PDP, slow decay.}
	 \label{fig:slow}
\end{figure}

The spectral efficiency for exponential PDP with slow decay is depicted in Fig. \ref{fig:slow}.
%
%\vspace{-.2cm}
In this case, the sub-optimal solutions achieves less than $93\%$ and $84\%$ of the achievable spectral efficiency of the optimal CIA precoder at high and low SNRs respectively. In general, both solutions suffer from a significant loss if compared to the uniform PDP case. This time, the less frequency selective channel resulting from the non uniform power distribution of the channel paths, diminishes the diversity and impacts negatively on the efficiency of the secondary link transmission. In particular, as the PDP departs form a uniform structure, a reduction on the amount of effective eigenmodes of the equivalent channel is seen, irrespective of the fact that number of transmit dimensions remain the same. This impacts the performance of the orthonormal root-based VFDM precoder as well. In fact, if the effective delay spread of the channel becomes shorter, the amount of non-zero roots of the channel diminishes. Moreover, their sparse power distribution yields a very ineffective orthonormalization process, resulting in a spectral efficiency loss for the secondary link, w.r.t. the CIA precoder, of as much as $25\%$ at high SNR.

\begin{figure}[!h]
	\centering
	\includegraphics[width=\linewidth]{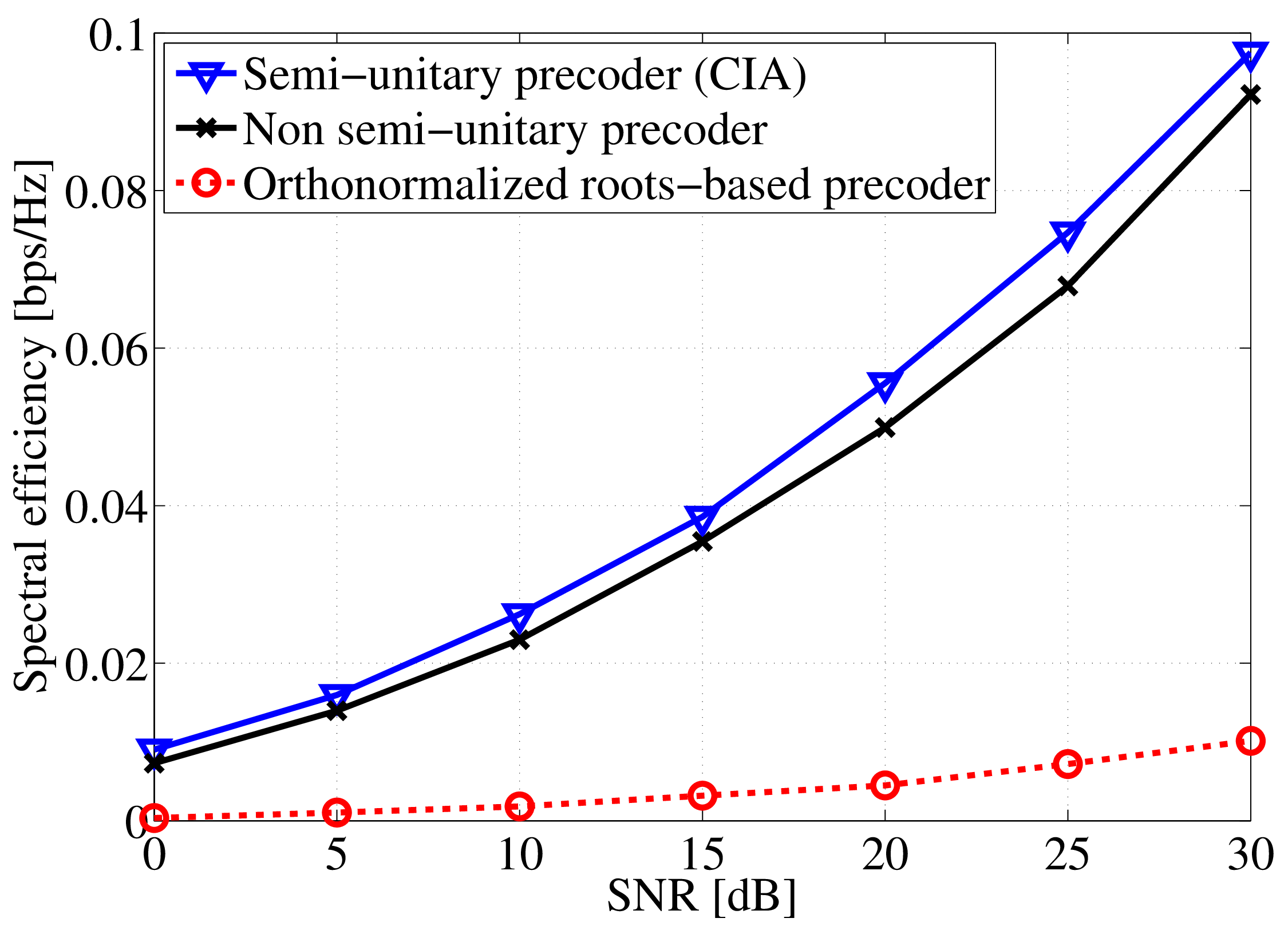} % requires the graphicx package
	\vspace{-0.8cm}
	\caption{Spectral efficiency of the secondary link. Exponential PDP, fast decay.}
	 \label{fig:fast}
\end{figure}

In Fig. \ref{fig:fast}, the spectral efficiency for exponential PDP with fast decay is shown.
%
%\vspace{-.2cm}
In this case, all techniques experience a considerable drop in spectral efficiency due to the very low amount of effective eigenmodes of the equivalent channel. Despite this, we note that the behavior of the optimal and sub-optimal strategy maintains a similar trend as in the previous case. Conversely, the orthonormal root-based VFDM precoder performance loss is of more than $90\%$, confirming the impact of the delay spread on the robustness of the root-based precoder computation. 

\vspace{-.2cm}
\section{Conclusion} \label{sec:conclusion}

In this work, a technique called cognitive interference alignment has been proposed to increase the spectral efficiency of the two-tiered network. This technique preserves the degrees of freedom of the legacy OFDM transmission, guaranteeing the presence of $N$ interference free dimensions at the MUE, while providing $L$ additional transmit dimensions to the SBS. The optimal linear strategy to maximize the spectral efficiency of the secondary link has been derived and tested for several channel models. We have shown that, for uniform PDP channels, the performance of the CIA and VFDM root-based precoder coincide, providing the optimal performance. For exponentially decaying PDP channels, CIA precoder shows a higher consistency w.r.t. the other considered approaches, outperforming both VFDM root-based and non optimally designed precoders. Nevertheless, the spectral efficiency of the secondary link highly hinges on the r.m.s. delay spread and PDP of the channel, and a greater frequency selectivity is preferable in terms of performance for the CIA scheme. The non-negligible loss induced by these channels opens a new research front, leading towards the design of suitable receiver architecture to compensate this spectral efficiency reduction. Techniques and algorithms to address this issue will be subject of future research, as well as the analysis of the impact of the primary transmission on the secondary link performance.

\vspace{-.2cm}
\begin{small}
\bibliographystyle{unsrt}
\bibliography{cia}
\end{small}

\end{document}